\documentclass[showpacs,aps,Numberfix,prx,reprint,superscriptaddress]{revtex4-1}

\usepackage{xfrac}
\usepackage{amssymb}
\usepackage{braket}
\usepackage{graphicx}
\usepackage{verbatim}
\usepackage{etoolbox}
\usepackage{amsfonts} 
\usepackage{amsmath} 
\usepackage[utf8]{inputenc} 
\usepackage{mathtools}
\usepackage{soul,xcolor,colortbl}
\usepackage{setspace}
\usepackage{multirow}
\usepackage{hyperref} \hypersetup{colorlinks=true,citecolor=blue,linkcolor=blue,urlcolor=blue} 
\usepackage{makecell}
\usepackage{bm} 
\usepackage{array}
\newcolumntype{C}[1]{>{\centering\arraybackslash}p{#1}}\usepackage{soul}
\definecolor{Gray}{gray}{0.85}

\usepackage{color} 
\usepackage{morefloats}

\definecolor{Gray}{gray}{0.9}
\definecolor{LightCyan}{rgb}{0.88,1,1}

\DeclareFixedFont{\ttb}{T1}{txtt}{bx}{n}{9} 
\DeclareFixedFont{\ttm}{T1}{txtt}{m}{n}{9}  

\definecolor{deepblue}{rgb}{0,0,0.5}
\definecolor{deepred}{rgb}{0.6,0,0}
\definecolor{deepgreen}{rgb}{0,0.5,0}
\definecolor{pranab_green}{rgb}{0,0.5,0}

\usepackage{listings}

\newcommand\pythonstyle{\lstset{
language=Python,
basicstyle=\ttm,
otherkeywords={self},             
keywordstyle=\ttb\color{deepblue},
emph={MyClass,__init__},          
emphstyle=\ttb\color{deepred},    
stringstyle=\color{deepgreen},
frame=tb,                         
breaklines=true,
showstringspaces=false            %
}}

\lstnewenvironment{python}[1][]
{
\pythonstyle
\lstset{#1}
}
{}

\newcommand\pythoninline[1]{{\pythonstyle\lstinline!#1!}}

\usepackage{comment}

\makeatletter \renewcommand\frontmatter@abstractwidth{\dimexpr\textwidth\relax} \makeatother  

\def\AFLOW{{\small AFLOW}}
\def\AFLOWPI{{\small AFLOW$\pi$}}
\def\AFLOWML{{\small AFLOW-ML}}
\def\JSON{{\small JSON}}
\def\POSCAR{{\small POSCAR}}
\def\POSCARFIVE{{\small POSCAR 5}}
\def\URL{{\small URL}}
\def\REST{{\small REST}}
\def\API{{\small API}}

\def\VASP{{\small VASP}}
\def\AEL{{\small AEL}}
\def\AGL{{\small AGL}}

\def\ICSD{{\small ICSD}}
\def\ML{{\small ML}}
\def\SDK{{\small SDK}}
\def\CLI{{\small CLI}}

\def\citeAFLOW{\cite{aflowPAPER,curtarolo:art110,curtarolo:art104, curtarolo:art63,curtarolo:art57,curtarolo:art49,monsterPGM,curtarolo:art127}}

\def\citeAFLOWLIB{\cite{aflowlibPAPER,curtarolo:art58,curtarolo:art92,curtarolo:art128}}

\begin{document}



\title{AFLOW-ML: A RESTful API for machine-learning predictions of materials properties}

\def\MEMS{\affiliation{Department of Mechanical Engineering and Materials Science, Duke University, Durham, North Carolina 27708, USA}}
\def\CMS{\affiliation{Center for Materials Genomics, Duke University,
 Durham, North Carolina 27708, USA}}
\def\FHI{\affiliation{Fritz-Haber-Institut der Max-Planck-Gesellschaft, 14195 Berlin-Dahlem, Germany}}

\author{Eric Gossett} \MEMS \CMS
\author{Cormac Toher} \MEMS \CMS
\author{Corey Oses} \MEMS \CMS
\author{Olexandr Isayev} 
\affiliation{Laboratory for Molecular Modeling, Division of Chemical Biology and Medicinal Chemistry, UNC Eshelman School of
 Pharmacy, University of North Carolina, Chapel Hill, North Carolina 27599, USA}
\author{Fleur Legrain}
\affiliation{LITEN, CEA-Grenoble, 38054 Grenoble, France}
\affiliation{Universite{\'e} Grenoble Alpes, 38000 Grenoble, France}
\author{Frisco Rose} \MEMS \CMS
\author{Eva Zurek} 
\affiliation{Department of Chemistry, State University of New York at Buffalo, Buffalo, New York 14260, USA}
\author{Jes{\'u}s Carrete} 
\affiliation{Institute of Materials Chemistry, TU Wien, A-1060 Vienna, Austria}
\author{Natalio Mingo} 
\affiliation{LITEN, CEA-Grenoble, 38054 Grenoble, France}
\author{Alexander Tropsha} \affiliation{Laboratory for Molecular Modeling, Division of Chemical Biology and Medicinal Chemistry, UNC Eshelman School of Pharmacy, University of North Carolina, Chapel Hill, North Carolina 27599, USA}
\author{Stefano Curtarolo} \email{stefano@duke.edu} \MEMS \CMS \FHI

\date{\today}

\begin{abstract}
\noindent Machine learning approaches, enabled by the emergence of comprehensive databases of materials properties, are becoming a fruitful direction for materials analysis.
As a result, a plethora of models have been constructed and trained on existing data to predict properties of new systems.
These powerful methods allow researchers to target studies only at interesting materials -- neglecting the non-synthesizable systems and those without the desired properties ---
thus reducing the amount of resources spent on expensive computations and/or time-consuming experimental synthesis.
However, using these predictive models is not always straightforward. 
Often, they require a panoply of technical expertise, creating barriers for general users.
\AFLOWML\ (\AFLOW\ \underline{M}achine \underline{L}earning) overcomes the problem by streamlining the use of the machine learning methods developed within the \AFLOW\ consortium.
The framework provides an open RESTful \API\ to directly access the continuously updated algorithms, 
which can be transparently integrated into any workflow to retrieve predictions of electronic, thermal and mechanical properties.
These types of interconnected cloud-based applications are envisioned
to be capable of further accelerating the adoption of machine learning methods into materials development.
\end{abstract}



\maketitle 

\section{Introduction}
\label{introduction}

Since their inception, high throughput materials science frameworks such as \AFLOW\ \citeAFLOW\ have been amassing large databases of materials properties. 
For instance, the \AFLOW\ database \citeAFLOWLIB\ alone contains over 1.7 million material compounds with over 170 million calculated
properties, generated from the \underline{I}norganic \underline{C}rystal \underline{S}tructure \underline{D}atabase (\ICSD) \cite{ICSD,ICSD0,Belsky_ActaCristB_2002}, as well as by decorating crystal structure prototypes \cite{curtarolo:art121}. 
Combined with other online databases such as the Materials Project \cite{materialsproject.org}, NoMaD \cite{nomad} and {\small OQMD} \cite{oqmd.org}, materials data is abundant and available. 
As a result, \underline{m}achine \underline{l}earning (\ML) methods have emerged as the ideal tool for 
data analysis \cite{Zeng_ChemMat_2002,Gu_SSS_2006,Ghiringhelli_PRL_2015,PyzerKnapp_AdFM_2015,Guzik_NMat_2016,Ziatdinov_NPJCM_MachineVision_2017},
by identifying the key features in a data set \cite{Michalski:1986:MLA:21934} to construct a model for predicting the properties of a material. 
Recently, several models were developed to predict the properties of various material classes such as perovskites \cite{curtarolo:art120,Pilania_SR_2016}, 
oxides \cite{Ceder_Chem_Materials}, elpasolites \cite{Faber_PRL_Elpasolite_2016,Pilania_CMS_Elpasolite_2017}, thermoelectrics \cite{curtarolo:art84,curtarolo:art85,Furmanchuk_JCC_ML_Seebeck_2017}, 
and metallic glasses \cite{acs.jpclett.7b01046}. 
Additionally, generalized approaches have been devised for inorganic materials \cite{curtarolo:art124,curtarolo:art129,Furmanchuk_RSCA_2016,Ward_ML_GFA_NPGCompMat_2016,deJong_SR_2016,Davies_Chem_Screening_2016,Reinhart_SM_ML_crystal_2017,curtarolo:art94}
and for systematically identifying efficient physical models of materials properties \cite{curtarolo:art135}.

While predictions are powerful tools for rational materials design, the discipline is still reasonably new within the realm of materials science. 
As a result, a working understanding of machine learning principles, along with a high level of technical expertise, is required for using code bases effectively. 
This inhibits accessibility, where an average end user aims to utilize the codes to retrieve predictions with as little complication as possible. 
With the ever increasing number of predictive models, a unique challenge emerges: 
how does one create an accessible means to integrate machine learning frameworks into a materials discovery workflow? 

\AFLOWML\ alleviates this issue by providing a simplified abstraction on top of sophisticated predictive models. 
Predictions are exposed through a web accessible \underline{A}pplication \underline{P}rogramming \underline{I}nterface (\API) where functionality is distilled down to its essence: from the user input, return a prediction. 
\AFLOWML\ can be added into any code base through use of an {\small HTTP} request library, native to most languages. 
Alternatively, \AFLOWML\ can be utilized though use of the included Python client and command line interface. 
Through such abstractions, \AFLOWML\ is accessible to a wide audience: it unburdens the user from having to understand the intricacies of machine learning and eliminates the technical expertise required to set up such codes.

\section{REST API}
\label{Features}

The \AFLOWML\ \API\ is structured around a \underline{RE}presentational \underline{S}tate \underline{T}ransfer (\REST) architecture, 
which allows resources to be accessed using {\small HTTP} request methods. 
Each resource is located at an endpoint, which is identified by a \URL\ comprised of descriptive nouns. 

A \URL\ may also include special variables in the form of path parameters and a query string. A path parameter is a segment of the \URL\ that specifies a particular resource and is denoted by a noun within braces (\{...\}) inside the endpoint. The query string is a list of key-value pairs placed at the end of a \URL, which control the representation (e.g. format) of the resource. Typically, it is used to apply filters or define the structure of the returned representation.

Resources within the \API\ are represented in \underline{J}ava\underline{S}cript \underline{O}bject \underline{N}otation (\JSON), and are referred to as objects. Once at an endpoint, the user must specify how to interact with the object. This is referred to as an action, and is an {\small HTTP} request method. The \API\ supports the two most common {\small HTTP} request methods, \verb|GET| and \verb|POST|, where \verb|GET| fetches an object and \verb|POST| sends user defined data to the server. Therefore, users will interact with the \API\ by performing actions (\verb|GET|, \verb|POST|) on endpoints (URLs) to retrieve objects (resources).

\begin{table}
 \begin{tabular}{|c|l|p{1.6cm}|}
  \hline
  \textbf{Action} & \textbf{Endpoint} & \textbf{Object} \\
  \hline
  \verb|POST| & \verb /plmf/prediction & task \\
  \verb|POST| & \verb /mfd/prediction & task \\
  \verb|GET| & \verb /prediction/result/{id} & status or prediction \\
  \hline
\end{tabular}
\caption{A list of all endpoints in the  \API. Actions specify the supported {\small HTTP} request method, endpoints define the \URL\ and objects are the returned resource. An endpoint with \{...\} denotes a path parameter.}
\label{table:endpoints}
\end{table}

\section{API Structure}

\begin{figure*}
  \centering
  \includegraphics[width=0.8\textwidth]{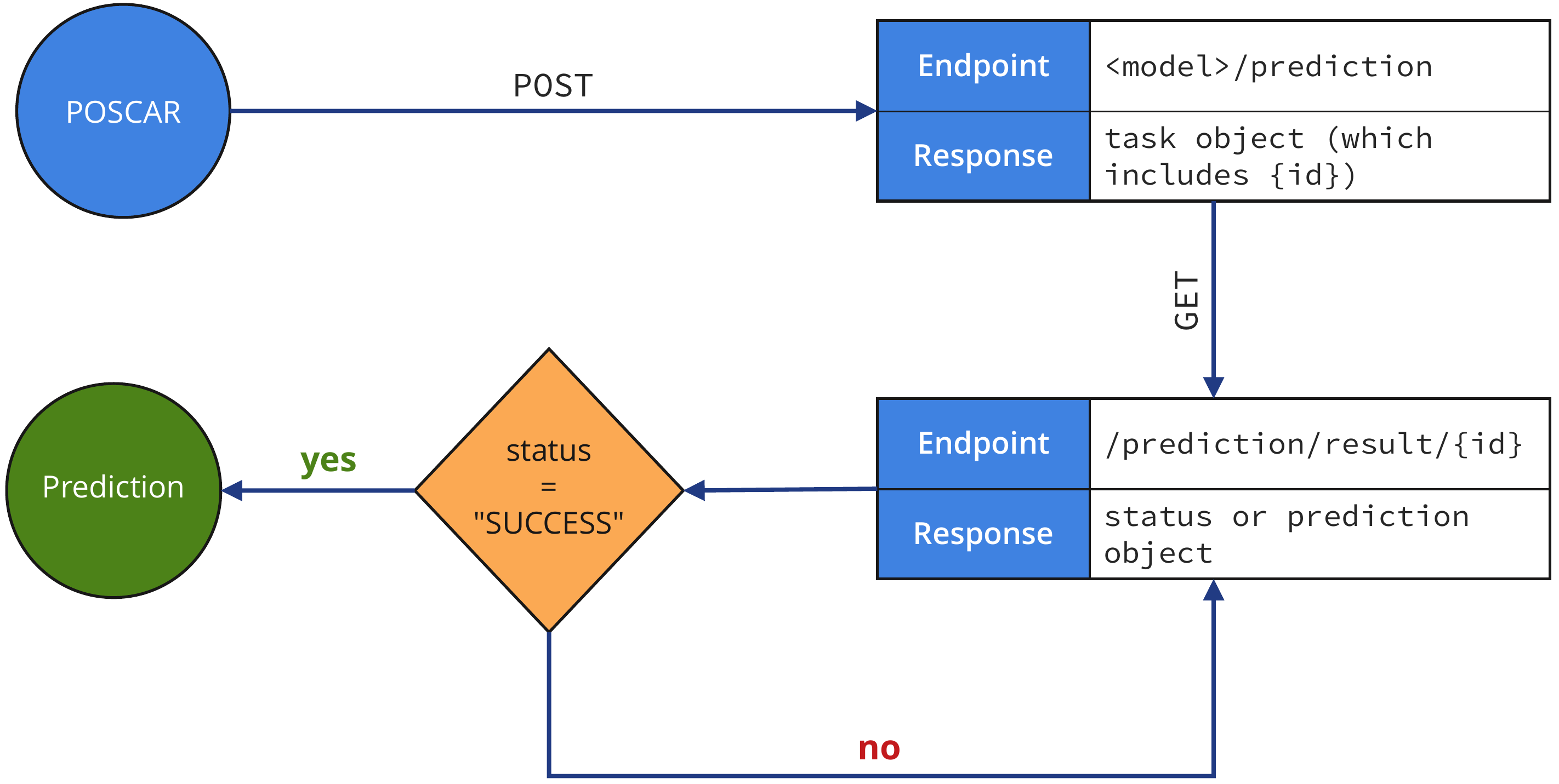}
  \caption{A flowchart illustrating the typical use-case of the API. First a \POSCAR\ file is posted to the \texttt{<model>/prediction} endpoint to retrieve a task object. The \texttt{id} field of the task object is used to poll the \texttt{<model>/prediction/\{id\}} for the status of the prediction. If the status field of the response equals \texttt{SUCCESS} then the prediction object is returned.}
  \label{fig:flowchart}
\end{figure*}

The overall structure of the \API\ can be seen in Table \ref{table:endpoints}. All endpoints are located at the base \URL\ \textbf{aflow.org/API/aflow-ml/v1.0/} and are organized by the model and the returned object.

\AFLOWML\ currently supports two models: \underline{m}olar \underline{f}raction \underline{d}escriptor (\verb|mfd|) \cite{curtarolo:art129} and \underline{p}roperty-\underline{l}abeled \underline{m}aterials \underline{f}ragments (\verb|plmf|) \cite{curtarolo:art124}. It is designed to be extensible, and additional models will be added in the future as they are developed. 

The \verb|mfd| model \cite{curtarolo:art129} predicts the material properties based on the chemical formula only: the vector of descriptors has 87 components, each component $b_{i}$ being the mole fraction of element $Z_{i}$ in the compound ($Z_{1}$ is H, $Z_{2}$ is He, etc.). The model is built with nonlinear support vector machines and a radial basis function kernel. The model is trained using a data set of 292 randomly selected compounds of the \ICSD\ for which the vibrational properties are computed with {\small DFT} calculations. 

The \verb|plmf| model \cite{curtarolo:art124} represents each crystal structure as a colored graph, where the atomic vertices are decorated by the reference properties of the corresponding elemental species. 
Topological neighbors are determined using a Voronoi tessellation, and these nodes are connected to form the graph. 
The final feature vector for the ML model is obtained by partitioning the full graph into smaller subgraphs, termed fragments in analogy with the fragment-based descriptors used in cheminformatics \cite{Ruggiu_MI_2010}. 
All \verb|plmf| models are built with the \underline{G}radient \underline{B}oosting \underline{D}ecision \underline{T}ree ({\small GBDT}) method \cite{Friedman_AnnStat_2001}. 
Models for electronic and thermo-mechanical properties were trained on 26,674 and 2,829 materials entries from the \AFLOW\ repository, respectively. 
All models are validated through $Y$-randomization (label scrambling) and five-fold cross validation. 

In general, \API\ usage involves uploading a material structure to a \verb|POST| endpoint, \verb|<model>/prediction|, and retrieving a prediction object from a \verb|GET| endpoint, \verb|/prediction/result/{id}|, as shown in the flowchart in Figure \ref{fig:flowchart}. \verb|POST| endpoints are responsible for the submission of a material structure for a prediction. In their request body, the file keyword is required. It must contain a string representation of the material's crystal structure, in \POSCARFIVE\ format (the lattice geometry and atomic position input file for version 5 of the \VASP\ {\small DFT} package \cite{kresse_vasp, vasp}). Upon receiving a request, the response body returns a task object containing information about the submitted structure, which has the following format:

\begin{verbatim}
{
  "id": String,
  "model": String,  
  "results_endpoint": String
}
\end{verbatim}

When a material is posted to the \API, a prediction task is created and added to a queue. Each task is assigned an identifier, the \verb|id| keyword, used to fetch the the prediction object at the endpoint referenced in the \verb|results_endpoint| keyword. This endpoint, \verb|/prediction/result/{id}|, supports the \verb|GET| method and requires the id as a path parameter. Depending on the status of the prediction task, the response body returns a status object or prediction object. When the task is still in the queue, the status object is returned:

\begin{verbatim}
{
  "status": String,
  "description": String
}
\end{verbatim}

The status object details the state of the prediction task. The state is identified by the keyword  \verb|status| which takes one of the following values: \verb|STARTED|, \verb|PENDING|, \verb|SUCCESS| and \verb|FAILURE|, while a description of each status type is given by the \verb|description| keyword. When attempting to retrieve the prediction object, it is best to poll the endpoint periodically to check the status. Tasks that are still within the queue are given the \verb|STARTED| or \verb|PENDING| status, while a completed task status will read \verb|SUCCESS|. In instances where the uploaded file is incorrectly formatted, a failed task will occur, status: \verb|FAILURE|. When the task is completed, the response contains the prediction object. The prediction object is an extension of the status object and contains different keywords depending on the model used. For \verb|plmf|, the prediction object, known as a \verb|plmf| prediction, takes the following form:

\begin{verbatim}
{
  "status": String,
  "description": String,
  "model": String,
  "citation": String,
  "ml_egap_type": String,
  "ml_egap": Number,
  "ml_energy_per_atom": Number,
  "ml_ael_bulk_modulus_vrh": Number,
  "ml_ael_shear_modulus_vrh": Number,
  "ml_agl_debye": Number,
  "ml_agl_heat_capacity_Cp_300K": Number,
  "ml_agl_heat_capacity_Cv_300K": Number,
  "ml_agl_heat_capacity_Cp_300K_per_atom": Number,
  "ml_agl_heat_capacity_Cv_300K_per_atom": Number,
  "ml_agl_thermal_conductivity_300K": Number,
  "ml_agl_thermal_expansion_300K": Number
}
\end{verbatim}

\noindent The \verb|mfd| model returns an \verb|mfd| prediction object:

\begin{verbatim}
  {
    "status": String,
    "description": String,
    "model": String,
    "citation": String,
    "ml_Cv": Number,
    "ml_Fvib": Number,
    "ml_Svib": Number
  }
  \end{verbatim}

\noindent The details of each object and their keywords are found in the List of \API\ Endpoints and Objects section.

\begin{widetext}

\section{Using the API}

The process to retrieve a prediction is as follows: First, the contents of a \POSCARFIVE\ file, titled test.poscar, are uploaded to the submission endpoint. This can be achieved by using an {\small HTTP} library such as requests (Python) \cite{pythonRequests}, 
URLSession (iOS \SDK), HttpURLConnection (Android \SDK), Fetch (JavaScript) or using a command line tool such as wGET or cURL. For this example, cURL will be used. The contents of the \POSCAR\ are posted to the submission endpoint as follows:

\begin{verbatim}
curl  http://aflow.org/API/aflow-ml/v1.0/plmf/prediction --data-urlencode file@test.poscar
\end{verbatim}

\noindent where the \verb|--data-urlencode| flag handles encoding the contents of the \POSCAR, located in the current directory, and associating it to the file keyword. Note that as mentioned previously, the \verb|POST| will return a \JSON\ response including the task id. The task id is then used to poll the results endpoint:

\begin{verbatim}
curl  http://aflow.org/API/aflow-ml/v1.0/prediction/result/{id}
\end{verbatim}

The status keyword is used as an indicator to determine if any additional polling is required. Depending on the status of the job, the endpoint will return either a task status object or a prediction object. If the status keyword's value is \verb|SUCCESS| then no additional polling is required, since the endpoint will return a prediction object, which is an extension of the task status object.

\end{widetext}

\section{List of API endpoints and objects}

This section includes the details of each endpoint and object accessible in the \API. Endpoint information contains the associated {\small HTTP} method, request parameters, request body data and response object for each endpoint. Object properties are listed along with their type and description.

\subsection{Endpoints}
\begin{itemize}
  
    \item \verb|POST  plmf/prediction|
      \begin{itemize}
        \item \textit{Description.} Uploads the contents of a \POSCARFIVE\ to retrieve a prediction using the \verb|plmf| model. 
        \item \textit{Query parameters.}
        \begin{itemize}
          \item \verb|file| (required) - The contents of the \POSCARFIVE.
        \end{itemize}
        \item \textit{Response format.} On success, the response header contains the {\small HTTP} status code 200 OK and the response body contains a task object, in \JSON\ format.
        \item \textit{Example.}
        \begin{verbatim}
curl http://aflow.org/API/aflow-ml/
v1.0/plmf/prediction
--data-urlencode file@test.poscar
          \end{verbatim}
      \end{itemize}

      \item \verb|POST  mfd/prediction|
      \begin{itemize}
        \item \textit{Description.} Uploads the contents of a \POSCARFIVE\ to retrieve a prediction using the \verb|mfd| model. 
        \item \textit{Query parameters.}
        \begin{itemize}
          \item \verb|file| (required) - The contents of the \POSCARFIVE.
        \end{itemize}
        \item \textit{Response format.} On success, the response header contains the {\small HTTP} status code 200 OK and the response body contains a task object, in \JSON\ format.
        \item \textit{Example.}
        \begin{verbatim}
curl http://aflow.org/API/aflow-ml/
v1.0/mfd/prediction
--data-urlencode file@test.poscar
          \end{verbatim}
      \end{itemize}

      \item \verb|GET  /prediction/result/{id}|
      \begin{itemize}
        \item \textit{Description.} Fetches the status object or returns the prediction object when the task is completed. 
        \item \textit{Path parameters.} 
        \begin{itemize}
          \item \verb|id| (required) - The unique identifier retrieved from the task object on submission. 
        \end{itemize}
        \item \textit{Query string arguments.} 
        \begin{itemize}
          \item \verb|fields| - A comma separated list of the fields to include in the \JSON\ response object. Note that specified fields will only affect the prediction object.
        \end{itemize}
        \item \textit{Response format.} On success, the response header contains the {\small HTTP} status code 200 OK. If the task is still pending, the response body contains a task status object in \JSON\ format. Upon completion the response body contains a prediction object in \JSON\  format.
      \end{itemize}
      \item \textit{Example.}
      \begin{verbatim}
curl http://aflow.org/API/aflow-ml/
v1.0/prediction/result
/59ea0f78-4868-4a1e-9a20-e7343f00907d
      \end{verbatim}

\end{itemize}

\subsection{Objects}
\begin{itemize}
  
    \item
      Task
      \begin{itemize}
        \item \textit{Description.} Describes the task for a submitted prediction. Includes the unique identifier for the prediction and results 
        endpoint. 
        \item \textit{Keys.} 
          \begin{itemize}
            \item \verb|id| 
              \begin{itemize}
                \item \textit{Description.} The unique identifier of the prediction task. Used at the fetch prediction endpoint to retrieve the status of a prediction and return the results on completion.
                \item \textit{Type.} String.
              \end{itemize}
            \item \verb|results_endpoint| 
              \begin{itemize}
                \item \textit{Description.} The path of the endpoint where the prediction task status and results are retrieved.
                \item \textit{Type.} String.
              \end{itemize}
            \item \verb|model| 
              \begin{itemize}
                \item \textit{Description.} The name of the machine learning model used to generate the prediction.
                \item \textit{Type.} String.
              \end{itemize}
          \end{itemize}
        \item \textit{Example.}
        \begin{verbatim}
{
    "id":     "d29af704-06bf
              -4dc8-8928
              -cd2c41aea454",
    "model":   "plmf",
    "results_endpoint": 
              "/prediction/result/
              d29af704-06bf
              -4dc8-8928-
              cd2c41aea454"
}
          \end{verbatim}
      \end{itemize}

      \item
      Status
      \begin{itemize}
        \item \textit{Description.} Provides the status of a (prediction) task.
        \item \textit{Keys.} 
          \begin{itemize}
            \item \verb|status| 
              \begin{itemize}
                \item \textit{Description.} The status of the task. Takes the following values: \verb|STARTED|, \verb|PENDING|, \verb|SUCCESS| and \verb|FAILURE|. When a task is added to the queue its status will read \verb|PENDING|. Once it reaches the top of the queue the status will read \verb|STARTED| and the prediction will run. If the prediction is successful the status will read \verb|SUCCESS|.
                \item \textit{Type.} String.
              \end{itemize}
            \item \verb|Description| 
              \begin{itemize}
                \item \textit{Description.} Describes the status of the task. 
                \item \textit{Type.} String.
              \end{itemize}
          \end{itemize}
        \item \textit{Example.}
        \begin{verbatim}
{
    "status":       "PENDING"  
    "description":  "The calculation
                     is running"
}
          \end{verbatim}
      \end{itemize}

      \item
      \verb|plmf| prediction
      \begin{itemize}
        \item \textit{Description.} The results of the prediction using the \verb|plmf| model. This is an extension of the task status object.
        \item \textit{Keys.} 
          \begin{itemize}
            \item \verb|status| 
              \begin{itemize}
                \item \textit{Description.} The status of the task. Takes the following values: \verb|STARTED|, \verb|PENDING|, \verb|SUCCESS| and \verb|FAILURE|. When a task is added to the queue its status will read \verb|PENDING|. Once it reaches the top of the queue the status will read \verb|STARTED| and the prediction will run. If the prediction is successful the status will read \verb|SUCCESS|.
                \item \textit{Type.} String.
              \end{itemize}
            \item \verb|description| 
              \begin{itemize}
                \item \textit{Description.} Describes the status of the task. 
                \item \textit{Type.} String.
              \end{itemize}
            \item \verb|model| 
            \begin{itemize}
              \item \textit{Description.} The model used in the prediction. 
              \item \textit{Type.} String.
            \end{itemize}
            \item \verb|citation| 
              \begin{itemize}
                \item \textit{Description.} The DOI for the model's publication.
                \item \textit{Type.} String.
              \end{itemize}
            \item \verb|ml_egap_type| 
              \begin{itemize}
                \item \textit{Description.} Specifies if the material is a metal or an insulator. Takes the following values: \verb|Metal|, \verb|Insulator|.
                \item \textit{Type.} String.
              \end{itemize}
            \item \verb|ml_egap| 
              \begin{itemize}
                \item \textit{Description.} The electronic band gap.
                \item \textit{Units.} eV.
                \item \textit{Type.} Number.
              \end{itemize}
            \item \verb|ml_energy_per_atom| 
              \begin{itemize}
                \item \textit{Description.} The energy per atom.
                \item \textit{Units.} eV/atom.
                \item \textit{Type.} Number.
              \end{itemize}
              \item \verb|ml_ael_bulk_modulus_vrh| 
              \begin{itemize}
                \item \textit{Description.} The bulk modulus, trained using the Automatic Elasticity Library (\AEL). \cite{curtarolo:art115}
                \item \textit{Units.} GPa.
                \item \textit{Type.} Number.
              \end{itemize}
            \item \verb|ml_ael_shear_modulus_vrh| 
              \begin{itemize}
                \item \textit{Description.} The shear modulus, trained using \AEL.
                \item \textit{Units.} GPa.
                \item \textit{Type.} Number.
              \end{itemize}
            \item \verb|ml_agl_debye| 
              \begin{itemize}
                \item \textit{Description.} The Debye temperature, trained using the Automatic GIBBS Library (\AGL) \cite{curtarolo:art96}.
                \item \textit{Units.} K.                
                \item \textit{Type.} Number.
              \end{itemize}
            \item \verb|ml_agl_heat_capacity_Cp_300K| 
              \begin{itemize}
                \item \textit{Description.} The heat capacity at 300K and constant pressure, trained using \AGL.
                \item \textit{Units.} $k_\mathrm{B}$/cell.
                \item \textit{Type.} Number.
              \end{itemize}
            \item \verb|ml_agl_heat_capacity_Cp_300K_per_atom| 
              \begin{itemize}
                \item \textit{Description.} The heat capacity per atom at 300K and constant pressure, trained using \AGL.
                \item \textit{Units.} $k_\mathrm{B}$/atom.
                \item \textit{Type.} Number.
              \end{itemize}
            \item \verb|ml_agl_heat_capacity_Cv_300K| 
              \begin{itemize}
                \item \textit{Description.} The heat capacity at 300K and constant volume, trained using \AGL.
                \item \textit{Units.} $k_\mathrm{B}$/cell.
                \item \textit{Type.} Number.
              \end{itemize}
            \item \verb|ml_agl_heat_capacity_Cv_300K_per_atom| 
              \begin{itemize}
                \item \textit{Description.} The heat capacity per atom at 300K and constant volume, trained using \AGL.
                \item \textit{Units.} $k_\mathrm{B}$/atom.
                \item \textit{Type.} Number.
              \end{itemize}
              \item \verb|ml_agl_thermal_conductivity_300K|  
              \begin{itemize}
                \item \textit{Description.} The lattice thermal conductivity at 300K, trained using \AGL.
                \item \textit{Units.} W/(m K).
                \item \textit{Type.} Number.
              \end{itemize}
            \item \verb|ml_agl_thermal_expansion_300K| 
              \begin{itemize}
                \item \textit{Description.} The thermal expansion coefficient at 300K, trained using \AGL.
                \item \textit{Units.} K$^{-1}$.
                \item \textit{Type.} Number.
              \end{itemize}
          \end{itemize}
        \item \textit{Example.}
        \begin{verbatim}
{
    "status":
                "SUCCESS"
    "description":  
                "The job has 
                 completed.",
    "model": "plmf",
    "citation": "10.1038/ncomms15679",                
    "ml_egap_type": 
                "Insulator",
    "ml_egap":
                0.923,
    "ml_energy_per_atom": 
                -5.760,
    "ml_ael_bulk_modulus_vrh":
                178.538,
    "ml_ael_shear_modulus_vrh": 
                140.121,
    "ml_agl_debye": 
                713.892,
    "ml_agl_heat_capacity_Cp_300K": 
                23.362,
    "ml_agl_heat_capacity_Cp_300K
    _per_atom": 
                2.333,
    "ml_agl_heat_capacity_Cv_300K": 
                22.625,
    "ml_agl_heat_capacity_Cv_300K
    _per_atom": 
                2.311,
    "ml_agl_thermal_conductivity
    _300K": 
                2.792,
    "ml_agl_thermal_expansion_300K": 
                6.093e-05
}
          \end{verbatim}
      \end{itemize}

      \item
      \verb|mfd| prediction
      \begin{itemize}
        \item \textit{Description.} The results of a prediction using the \verb|mfd| model. This is an extension of the task status object.
        \item \textit{Keys.} 
          \begin{itemize}
            \item \verb|status| 
              \begin{itemize}
                \item \textit{Description.} The status of the task. Takes the following values: \verb|STARTED|, \verb|PENDING|, \verb|SUCCESS| and \verb|FAILURE|. When a task is added to the queue its status will read \verb|PENDING|. Once it reaches the top of the queue the status will read \verb|STARTED| and the prediction will run. If the prediction is successful the status will read \verb|SUCCESS|.
                \item \textit{Type.} String.
              \end{itemize}
            \item \verb|description| 
              \begin{itemize}
                \item \textit{Description.} Describes the status of the task. 
                \item \textit{Type.} String.
              \end{itemize}
            \item \verb|model| 
            \begin{itemize}
              \item \textit{Description.} The model used in the prediction. 
              \item \textit{Type.} String.
            \end{itemize}
            \item \verb|citation| 
              \begin{itemize}
                \item \textit{Description.} The DOI for the model's publication.
                \item \textit{Type.} String.
              \end{itemize}
            \item \verb|ml_Cv| 
              \begin{itemize}
                \item \textit{Description.} The heat capacity at constant volume.
                \item \textit{Units.} meV/(atom K).
                \item \textit{Type.} Number.
              \end{itemize}
            \item \verb|ml_Fvib| 
              \begin{itemize}
                \item \textit{Description.} The vibrational free energy per atom. 
                \item \textit{Units.} meV/atom.
                \item \textit{Type.} Number.
              \end{itemize}
            \item \verb|ml_Svib| 
              \begin{itemize}
                \item \textit{Description.} The vibrational entropy per atom.
                \item \textit{Units.} meV/(atom K).
                \item \textit{Type.} Number.
              \end{itemize}
          \end{itemize}
        \item \textit{Example.}
        \begin{verbatim}
{
    "description": 
        "The job has completed.",
    "model": "mfd",
    "citation": 
         "10.1021/acs.chemmater.7b00789",
    "status": "SUCCESS",  
    "ml_Cv": 0.221,
    "ml_Fvib": 21.188,
    "ml_Svib": 0.211
}
          \end{verbatim}
      \end{itemize}

\end{itemize}

\section{Python Client}
\label{python-client}

\begin{figure}[h!]
  \centering
  \includegraphics[width=0.5\textwidth]{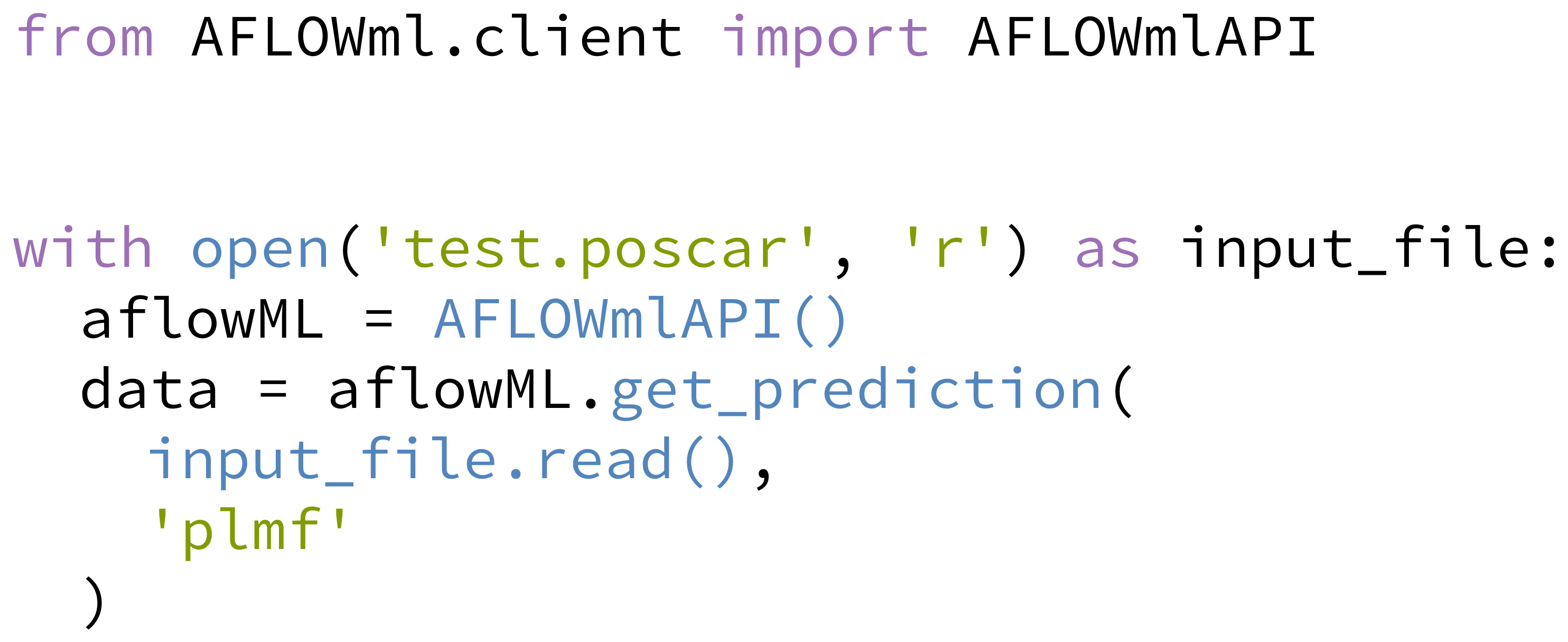}
  \caption{Example showing how to retrieve a prediction using the \AFLOWML\ Python client.}
\label{aflowmlclientcall}
\end{figure}

A Python client is available for the \AFLOWML\ \REST\ \API\ that provides researchers and developers a means to integrate \AFLOWML\ into their applications or workflows, such as \AFLOWPI\ \cite{curtarolo:art127}. 
The client includes the \verb|AFLOWmlAPI| class which provides all the
functionality needed to interface with the \AFLOWML\ \API, and can be downloaded at \verb|aflow.org/src/aflow-ml|. 
From the client, a prediction is retrieved by passing the contents of a \POSCAR\ to the \verb|get_prediction| method. 
The \verb|AFLOWmlAPI| class can be incorporated into a Python framework using code similar to the example illustrated in Figure \ref{aflowmlclientcall}.

This method takes two arguments: \verb|poscar| and \verb|model|, where \verb|poscar| is a file object (reading from the file \verb|test.poscar| in the example in Figure \ref{aflowmlclientcall}) and \verb|model| is a string specifying the model to use (\verb|plmf| or \verb|mfd|). 
This method returns a Python dictionary, in which the keys and respective predicted values are model dependent. For a list of each prediction object's key and value pair, please refer to the previous section.

The client's \verb|AFLOWmlAPI| class includes two additional methods, \verb|submit_job| and \verb|poll_job|, that provide more control when submitting a prediction, and which can be used in place of the \verb|get_prediction| method in the example shown in Figure \ref{aflowmlclientcall}. 
The \verb|submit_job| method targets the \verb|<model>/prediction| endpoint and returns the jobs task id. 
From this id, the job can be polled using the \verb|poll_job| method which will return a prediction object upon completion. 
These methods are ideal for cases where the user would prefer to postpone polling to a later time.

\section{Command Line Interface}

Upon installation, the Python \AFLOWML\ client provides a \underline{c}ommand \underline{l}ine \underline{i}nterface (\CLI) titled \verb|aflow-ml|. The \CLI\ exposes all the functionality of the Python client and is targeted at users who are not familiar with Python or using \REST\ APIs. To receive a prediction the path of the \POSCARFIVE\ file is passed to the \CLI\ as a positional argument. Additionally, the model type is specified via the \verb|--model| flag: 

\begin{verbatim}
aflow-ml  test.poscar --model=plmf
\end{verbatim}

By default, the \CLI\ will output the results to the terminal. The default functionality is modified by the use of additional flags. For instance, results can be saved to an out file by use of the \verb|-s| flag:

\begin{verbatim}
aflow-ml  test.poscar --model=plmf -s
\end{verbatim}

\noindent
where the predicted results are saved to a file titled \verb|prediction.txt|. Additional flags exist which provide various levels of customization such as specifying the predicted values to return or the format of the output. A list of each flag is found below. This list is also viewable from the \CLI\ using the \verb|-h| or \verb|--help| flags.

\subsection{\CLI\ flags}
\label{CLI-flags}

\begin{itemize}
  \item 
  Model
  \begin{itemize}
    \item \textit{Flag.} \verb|-m| or \verb|--model|
    \item \textit{Description.} (Required) Specifies the model to use in the prediction. 
    \item \textit{Example.} \\
    \verb|aflow-ml test.poscar --model=plmf|
  \end{itemize}

  \item
    Save
    \begin{itemize}
      \item \textit{Flag.} \verb|-s| or \verb|--save|
      \item \textit{Description.} Saves the prediction to a file. If the out file is not specified contents are saved to a file named \verb|prediction.txt|.
      \item \textit{Example.} \\
      \verb|aflow-ml test.poscar -m plmf -s|
    \end{itemize}

    \item
      Outfile
      \begin{itemize}
        \item \textit{Flag.} \verb|--format|
        \item \textit{Description.} Specifies the path and name of the out file.
        \item \textit{Example.} \\
        \verb|aflow-ml test.poscar -m plmf -s|
        \verb|--outfile=prediction.txt|
      \end{itemize}

    \item
      Format
      \begin{itemize}
        \item \textit{Flag.} \verb|--format|
        \item \textit{Description.} Specifies the format of the out file. Currently, text and \JSON\ are supported.
        \item \textit{Example.} \\
        \verb|aflow-ml test.poscar -m plmf -s|
        \verb| --format=JSON|
      \end{itemize}

      \item
        Fields
        \begin{itemize}
          \item \textit{Flag.} \verb|--fields|
          \item \textit{Description.} State the predicted fields to show in the output. Expects fields as a comma separated list. If the flag is not present, all fields will be shown.
          \item \textit{Example.} \\
          \verb|aflow-ml test.poscar -m plmf -s|
          \verb| --fields=egap_ml,egap_type_ml|
        \end{itemize}

        \item
          Verbose
          \begin{itemize}
            \item \textit{Flag.} \verb|-v| or \verb|--verbose|
            \item \textit{Description.} Toggle verbose mode. When enabled the \CLI\ will log the progress of the prediction.
            \item \textit{Example.} \\
            \verb|aflow-ml test.poscar -m plmf -v|
          \end{itemize}
\end{itemize}

\section{Conclusion}

\AFLOWML\ enhances materials discovery by providing streamlined open access to predictive models. 
The \REST\ \API\ promotes resource sharing, where any application, workflow or website may leverage our models. 
Additionally, the Python client provides a closed solution, which requires little programming knowledge to get started. 
With this flexibility, \AFLOWML\ presents the accessible option for machine learning in the materials design community.

\section*{Acknowledgments}

This work is partially supported by DOD-ONR (N00014-14-1-0526, N00014-16-1-2326, N00014-16-1-2583, N00014-17-1-2090), 
and by the Duke University---Center for Materials Genomics.
C.O. acknowledges support from the National Science Foundation Graduate Research Fellowship under Grant No. DGF-1106401.
A.T. and O.I. acknowledge support from DOD-ONR N00014-16-1-2311 and Eshelman Institute for Innovation Award. O.I. acknowledges Extreme Science and Engineering Discovery Environment (XSEDE) award DMR-110088, which is supported by National Science Foundation grant number ACI-1053575.
S.C. acknowledges the Alexander von Humboldt Foundation for financial support.


\section*{References}
\small
\begin{thebibliography}{10}

\bibitem{aflowPAPER}
S.~Curtarolo, W.~Setyawan, G.~L.~W. Hart, M.~Jahn\'{a}tek, R.~V. Chepulskii,
  R.~H. Taylor, S.~Wang, J.~Xue, K.~Yang, O.~Levy, M.~J. Mehl, H.~T. Stokes,
  D.~O. Demchenko, and D.~Morgan, \emph{{AFLOW}: An automatic framework for
  high-throughput materials discovery}, Comput.\ Mater.\ Sci. \textbf{58},
  218--226 (2012).

\bibitem{curtarolo:art110}
K.~Yang, C.~Oses, and S.~Curtarolo, \emph{Modeling Off-Stoichiometry Materials
  with a High-Throughput {\it Ab-Initio} Approach}, Chem.\ Mater. \textbf{28},
  6484--6492 (2016).

\bibitem{curtarolo:art104}
C.~E. Calderon, J.~J. Plata, C.~Toher, C.~Oses, O.~Levy, M.~Fornari, A.~Naturen,
  M.~J. Mehl, G.~L.~W. Hart, M.~{Buongiorno~Nardelli}, and S.~Curtarolo,
  \emph{The {AFLOW} standard for high-throughput materials science
  calculations}, Comput.\ Mater.\ Sci. \textbf{108 Part A}, 233--238 (2015).

\bibitem{curtarolo:art63}
O.~Levy, M.~Jahn\'{a}tek, R.~V. Chepulskii, G.~L.~W. Hart, and S.~Curtarolo,
  \emph{Ordered Structures in {Rh}enium Binary Alloys from First-Principles
  Calculations}, J.\ Am.\ Chem.\ Soc. \textbf{133}, 158--163 (2011).

\bibitem{curtarolo:art57}
O.~Levy, G.~L.~W. Hart, and S.~Curtarolo, \emph{Structure maps for {h}cp metals
  from first-principles calculations}, Phys.\ Rev.\ B \textbf{81}, 174106
  (2010).

\bibitem{curtarolo:art49}
O.~Levy, G.~L.~W. Hart, and S.~Curtarolo, \emph{Uncovering Compounds by Synergy
  of Cluster Expansion and High-Throughput Methods}, J.\ Am.\ Chem.\ Soc.
  \textbf{132}, 4830--4833 (2010).

\bibitem{monsterPGM}
G.~L.~W. Hart, S.~Curtarolo, T.~B. Massalski, and O.~Levy, \emph{Comprehensive
  Search for New Phases and Compounds in Binary Alloy Systems Based on
  {P}latinum-Group Metals, Using a Computational First-Principles Approach},
  Phys.\ Rev.\ X \textbf{3}, 041035 (2013).

\bibitem{curtarolo:art127}
A.~R. Supka, T.~E. Lyons, L.~S.~I. Liyanage, P.~{D'{A}mico},
  R.~{Al~Rahal~Al~Orabi}, S.~Mahatara, P.~Gopal, C.~Toher, D.~Ceresoli,
  A.~Calzolari, S.~Curtarolo, M.~{Buongiorno~Nardelli}, and M.~Fornari,
  \emph{{\small AFLOW}$\pi$: A minimalist approach to high-throughput {\it ab
  initio} calculations including the generation of tight-binding hamiltonians},
  Comput.\ Mater.\ Sci. \textbf{136}, 76--84 (2017).

\bibitem{aflowlibPAPER}
S.~Curtarolo, W.~Setyawan, S.~Wang, J.~Xue, K.~Yang, R.~H. Taylor, L.~J.
  Nelson, G.~L.~W. Hart, S.~Sanvito, M.~{Buongiorno~Nardelli}, N.~Mingo, and
  O.~Levy, \emph{{AFLOWLIB.ORG}: A distributed materials properties repository
  from high-throughput {\it ab initio} calculations}, Comput.\ Mater.\ Sci.
  \textbf{58}, 227--235 (2012).

\bibitem{curtarolo:art58}
W.~Setyawan and S.~Curtarolo, \emph{High-throughput electronic band structure
  calculations: Challenges and tools}, Comput.\ Mater.\ Sci. \textbf{49},
  299--312 (2010).

\bibitem{curtarolo:art92}
R.~H. Taylor, F.~Rose, C.~Toher, O.~Levy, K.~Yang, M.~{Buongiorno~Nardelli},
  and S.~Curtarolo, \emph{A {REST}ful {API} for exchanging materials data in
  the {AFLOWLIB}.org consortium}, Comput.\ Mater.\ Sci. \textbf{93}, 178--192
  (2014).

\bibitem{curtarolo:art128}
F.~Rose, C.~Toher, E.~Gossett, C.~Oses, M.~{Buongiorno~Nardelli}, M.~Fornari,
  and S.~Curtarolo, \emph{{AFLUX}: The {LUX} materials search {API} for the
  {AFLOW} data repositories}, Comput.\ Mater.\ Sci. \textbf{137}, 362--370
  (2017).

\bibitem{ICSD}
G.~Bergerhoff, R.~Hundt, R.~Sievers, and I.~D. Brown, \emph{The inorganic
  crystal structure data base}, J.\ Chem.\ Inf.\ Comput.\ Sci. \textbf{23},
  66--69 (1983).

\bibitem{ICSD0}
A.~D. Mighell and V.~L. Karen, \emph{NIST Materials Science Databases}, Acta\
  Crystallogr.\ Sect.\ A \textbf{49}, c409 (1993).

\bibitem{Belsky_ActaCristB_2002}
A.~Belsky, M.~Hellenbrandt, V.~L. Karen, and P.~Luksch, \emph{New developments
  in the {In}organic {C}rystal {S}tructure {D}atabase ({ICSD}): accessibility
  in support of materials research and design}, Acta\ Crystallogr.\ Sect.\ B
  \textbf{58}, 364--369 (2002).

\bibitem{curtarolo:art121}
M.~J. Mehl, D.~Hicks, C.~Toher, O.~Levy, R.~M. Hanson, G.~L.~W. Hart, and
  S.~Curtarolo, \emph{The {AFLOW} Library of Crystallographic Prototypes: Part
  1}, Comput.\ Mater.\ Sci. \textbf{136}, S1--S828 (2017).

\bibitem{materialsproject.org}
A.~Jain, G.~Hautier, C.~J. Moore, S.~P. Ong, C.~C. Fischer, T.~Mueller, K.~A.
  Persson, and G.~Ceder, \emph{A high-throughput infrastructure for density
  functional theory calculations}, Comput.\ Mater.\ Sci. \textbf{50},
  2295--2310 (2011).

\bibitem{nomad}
M.~Scheffler, C.~Draxl, and {Computer Center of the Max-Planck Society,
  Garching}, \emph{The {NoMaD} Repository}, http://nomad-repository.eu (2014).

\bibitem{oqmd.org}
J.~E. Saal, S.~Kirklin, M.~Aykol, B.~Meredig, and C.~Wolverton, \emph{Materials
  Design and Discovery with High-Throughput Density Functional Theory: The
  {O}pen {Q}uantum {M}aterials {D}atabase ({OQMD})}, JOM \textbf{65},
  1501--1509 (2013).

\bibitem{Zeng_ChemMat_2002}
Y.~Zeng, S.~J. Chua, and P.~Wu, \emph{On the Prediction of Ternary
  Semiconductor Properties by Artificial Intelligence Methods}, Chem.\ Mater.
  \textbf{14}, 2989--2998 (2002).

\bibitem{Gu_SSS_2006}
T.~Gu, W.~Lu, X.~Bao, and N.~Chen, \emph{Using support vector regression for
  the prediction of the band gap and melting point of binary and ternary
  compound semiconductors}, Solid\ State\ Sci. \textbf{8}, 129--136 (2006).

\bibitem{Ghiringhelli_PRL_2015}
L.~M. Ghiringhelli, J.~Vybiral, S.~V. Levchenko, C.~Draxl, and M.~Scheffler,
  \emph{Big Data of Materials Science: Critical Role of the Descriptor}, Phys.\
  Rev.\ Lett. \textbf{114}, 105503 (2015).

\bibitem{PyzerKnapp_AdFM_2015}
E.~O. Pyzer-Knapp, K.~{Li}, and A.~Aspuru-Guzik, \emph{Learning from the
  {H}arvard {Cl}ean {E}nergy {Pr}oject: The Use of Neural Networks to
  Accelerate Materials Discovery}, Adv.\ Func.\ Mater. \textbf{25}, 6495--6502
  (2015).

\bibitem{Guzik_NMat_2016}
R.~G\'{o}mez-Bombarelli, J.~Aguilera-Iparraguirre, T.~D. Hirzel, D.~Duvenaud,
  D.~Maclaurin, M.~A. Blood-Forsythe, H.~S. Chae, M.~Einzinger, D.-G. Ha,
  T.~Wu, G.~Markopoulos, S.~Jeon, H.~Kang, H.~Miyazaki, M.~Numata, S.~Kim,
  W.~Huang, S.~I. Hong, M.~Baldo, R.~P. Adams, and A.~Aspuru-Guzik,
  \emph{Design of efficient molecular organic light-emitting diodes by a
  high-throughput virtual screening and experimental approach}, Nature\ Mater.
  \textbf{15}, 1120--1127 (2016).

\bibitem{Ziatdinov_NPJCM_MachineVision_2017}
M.~Ziatdinov, A.~Maksov, and S.~V. Kalinin, \emph{Learning surface molecular
  structures via machine vision}, NPJ\ Comput.\ Mater. \textbf{3}, 31 (2017).

\bibitem{Michalski:1986:MLA:21934}
S.~R. Michalski, G.~J. Carbonell, and M.~T. Mitchell, eds., \emph{Machine
  Learning an Artificial Intelligence Approach Volume II} (Morgan Kaufmann
  Publishers Inc., San Francisco, CA, USA, 1986).

\bibitem{curtarolo:art120}
A.~{van~Roekeghem}, J.~Carrete, C.~Oses, S.~Curtarolo, and N.~Mingo,
  \emph{High-Throughput Computation of Thermal Conductivity of High-Temperature
  Solid Phases: The Case of Oxide and Fluoride Perovskites}, Phys.\ Rev.\ X
  \textbf{6}, 041061 (2016).

\bibitem{Pilania_SR_2016}
G.~Pilania, A.~Mannodi-Kanakkithodi, B.~P. Uberuaga, R.~Ramprasad, J.~E.
  Gubernatis, and T.~Lookman, \emph{Machine learning bandgaps of double
  perovskites}, Sci.\ Rep. \textbf{6}, 19375 (2016).

\bibitem{Ceder_Chem_Materials}
G.~Hautier, C.~C. Fischer, A.~Jain, T.~Mueller, and G.~Ceder, \emph{Finding
  Naturere'{s} Missing Ternary Oxide Compounds using Machine Learning and Density
  Functional Theory}, Chem.\ Mater. \textbf{22}, 3762--3767 (2010).

\bibitem{Faber_PRL_Elpasolite_2016}
F.~A. Faber, A.~Lindmaa, O.~A. von Lilienfeld, and R.~Armiento, \emph{Machine
  Learning Energies of 2 Million Elpasolite ({ABC}$_{2}${D}$_{6}$) Crystals},
  Phys.\ Rev.\ Lett. \textbf{117}, 135502 (2016).

\bibitem{Pilania_CMS_Elpasolite_2017}
G.~Pilania, J.~E. Gubernatis, and T.~Lookman, \emph{Multi-fidelity machine
  learning models for accurate bandgap predictions of solids}, Comput.\ Mater.\
  Sci. \textbf{129}, 156--163 (2017).

\bibitem{curtarolo:art84}
J.~Carrete, W.~{Li}, N.~Mingo, S.~Wang, and S.~Curtarolo, \emph{Finding
  Unprecedentedly Low-Thermal-Conductivity Half-Heusler Semiconductors via
  High-Throughput Materials Modeling}, Phys.\ Rev.\ X \textbf{4}, 011019
  (2014).

\bibitem{curtarolo:art85}
J.~Carrete, N.~Mingo, S.~Wang, and S.~Curtarolo, \emph{Nanograined
  Half-{H}eusler Semiconductors as Advanced Thermoelectrics: An Ab Initio
  High-Throughput Statistical Study}, Adv.\ Func.\ Mater. \textbf{24},
  7427--7432 (2014).

\bibitem{Furmanchuk_JCC_ML_Seebeck_2017}
A.~Furmanchuk, J.~E. Saal, J.~W. Doak, G.~B. Olson, A.~Choudhary, and
  A.~Agrawal, \emph{Prediction of seebeck coefficient for compounds without
  restriction to fixed stoichiometry: A machine learning approach}, J.\
  Comput.\ Chem.  (2017).

\bibitem{acs.jpclett.7b01046}
Y.~T. Sun, H.~Y. Bai, M.~Z. Li, and W.~H. Wang, \emph{Machine Learning Approach
  for Prediction and Understanding of Glass-Forming Ability}, J.\ Phys.\ Chem.\
  Lett. \textbf{8}, 3434--3439 (2017).

\bibitem{curtarolo:art124}
O.~Isayev, C.~Oses, C.~Toher, E.~Gossett, S.~Curtarolo, and A.~Tropsha,
  \emph{Universal fragment descriptors for predicting electronic properties of
  inorganic crystals}, Nature\ Commun. \textbf{8}, 15679 (2017).

\bibitem{curtarolo:art129}
F.~Legrain, J.~Carrete, A.~{van~Roekeghem}, S.~Curtarolo, and N.~Mingo,
  \emph{How Chemical Composition Alone Can Predict Vibrational Free Energies
  and Entropies of Solids}, Chem.\ Mater. \textbf{29}, 6220--6227 (2017).

\bibitem{Furmanchuk_RSCA_2016}
A.~Furmanchuk, A.~Agrawal, and A.~Choudhary, \emph{Predictive analytics for
  crystalline materials: bulk modulus}, RSC\ Adv. \textbf{6}, 95246--95251
  (2016).

\bibitem{Ward_ML_GFA_NPGCompMat_2016}
L.~Ward, A.~Agrawal, A.~Choudhary, and C.~Wolverton, \emph{A general-purpose
  machine learning framework for predicting properties of inorganic materials},
  NPJ\ Comput.\ Mater. \textbf{2}, 16028 (2016).

\bibitem{deJong_SR_2016}
M.~{de~Jong}, W.~Chen, R.~Notestine, K.~A. Persson, G.~Ceder, A.~Jain, M.~D.
  Asta, and A.~Gamst, \emph{A Statistical Learning Framework for Materials
  Science: Application to Elastic Moduli of $k$-nary Inorganic Polycrystalline
  Compounds}, Sci.\ Rep. \textbf{6}, 34256 (2016).

\bibitem{Davies_Chem_Screening_2016}
D.~W. Davies, K.~T. Butler, A.~J. Jackson, A.~Morris, J.~M. Frost, J.~M.
  Skelton, and A.~Walsh, \emph{Computational Screening of All Stoichiometric
  Inorganic Materials}, Chem \textbf{1}, 617--627 (2016).

\bibitem{Reinhart_SM_ML_crystal_2017}
W.~F. Reinhart, A.~W. Long, M.~P. Howard, A.~L. Ferguson, and A.~Z.
  Panagiotopoulos, \emph{Machine learning for autonomous crystal structure
  identification}, Soft Matter \textbf{13}, 4733--4745 (2017).

\bibitem{curtarolo:art94}
O.~Isayev, D.~Fourches, E.~N. Muratov, C.~Oses, K.~Rasch, A.~Tropsha, and
  S.~Curtarolo, \emph{Materials Cartography: Representing and Mining Materials
  Space Using Structural and Electronic Fingerprints}, Chem.\ Mater.
  \textbf{27}, 735--743 (2015).

\bibitem{curtarolo:art135}
R.~Ouyang, S.~Curtarolo, E.~Ahmetcik, M.~Scheffler, and L.~M. Ghiringhelli,
  \emph{SISSO: a compressed-sensing method for systematically identifying
  efficient physical models of materials properties}, submitted
  arxiv.org/1710.03319  (2017).

\bibitem{Ruggiu_MI_2010}
F.~Ruggiu, G.~Marcou, A.~Varnek, and D.~Horvath, \emph{{ISIDA}
  Property-Labelled Fragment Descriptors}, Mol.\ Informatics \textbf{29},
  855--868 (2010).

\bibitem{Friedman_AnnStat_2001}
J.~H. Friedman, \emph{Greedy Function Approximation: A Gradient Boosting
  Machine}, Ann.\ Stat. \textbf{29}, 1189--1232 (2001).

\bibitem{kresse_vasp}
G.~Kresse and J.~Hafner, \emph{{\it Ab initio} molecular dynamics for liquid
  metals}, Phys.\ Rev.\ B \textbf{47}, 558--561 (1993).

\bibitem{vasp}
G.~Kresse and J.~Furthm\"{u}ller, \emph{Efficient iterative schemes for {\it ab
  initio} total-energy calculations using a plane-wave basis set}, Phys.\ Rev.\
  B \textbf{54}, 11169--11186 (1996).

\bibitem{pythonRequests}
K.~Reitz, \emph{Python Requests Library},
  \url{http://docs.python-requests.org/en/master/}.

\bibitem{curtarolo:art115}
C.~Toher, C.~Oses, J.~J. Plata, D.~Hicks, F.~Rose, O.~Levy, M.~{de Jong}, M.~D.
  Asta, M.~Fornari, M.~{Buongiorno~Nardelli}, and S.~Curtarolo, \emph{Combining
  the {AFLOW} {GIBBS} and Elastic Libraries to efficiently and robustly screen
  thermomechanical properties of solids}, Phys.\ Rev.\ Mater. \textbf{1},
  015401 (2017).

\bibitem{curtarolo:art96}
C.~Toher, J.~J. Plata, O.~Levy, M.~{de~Jong}, M.~D. Asta,
  M.~{Buongiorno~Nardelli}, and S.~Curtarolo, \emph{High-throughput
  computational screening of thermal conductivity, {D}ebye temperature, and
  {G}r\"{u}neisen parameter using a quasiharmonic {D}ebye model}, Phys.\ Rev.\
  B \textbf{90}, 174107 (2014).

\end{thebibliography}
\end{document}